\documentclass[12pt,tightenlines,eqsecnum,floats,showpacs,nofootinbib,amsmath,amssymb,aps,prd]{revtex4}

\usepackage{graphicx}
\usepackage{amsmath,amssymb}
\usepackage{subfigure}

\def\be{\begin{equation}}
\def\ee{\end{equation}}
\def\ba{\begin{eqnarray}}
\def\ea{\end{eqnarray}}

\def\f{\frac}
\def\dd{{\rm d}}
\def\sgn{\mathrm{sgn}}
\def\l{\lambda}
\def\g{{\rm grav}}
\def\L{\mathcal{L}}
\def\ta{\tilde{\alpha}}
\def\tc{\tilde{c}}
\def\tp{\tilde{p}}
\def\e{{}^o\!e}
\def\w{{}^o\!\omega}
\def\q{{}^o\!q}
\def\xiz{{}^o\xi}
\def\C{{}^o C}
\def\b{{}^o b}
\def\lp{\ell_{\mathrm{Pl}}}
\def\m{{\rm matt}}
\def\ve{\varepsilon}
\def\R{\mathbb{R}}
\def\Z{\mathbb{Z}}
\def\H{\mathcal{H}}
\def\Hkg{\mathcal{H}_{\rm kin}^{\rm grav}}

\begin{document}


\title{Loop quantum cosmology of Bianchi type II models}

\author{Abhay Ashtekar}
\email{ashtekar@gravity.psu.edu}
\author{Edward Wilson-Ewing}
\email{wilsonewing@gravity.psu.edu} \affiliation{Institute for
Gravitation and the Cosmos, and Physics Department,\\ The
Pennsylvania State University, University Park, PA 16802, USA}

\begin{abstract}

The ``improved dynamics'' of loop quantum cosmology is extended to
include the Bianchi type II model. Because these space-times admit
both anisotropies and non-zero spatial curvature, certain technical
difficulties arise over and above those encountered in the analysis
of the (anisotropic but spatially flat) Bianchi type I space-times,
and of the (spatially curved but isotropic) k=$\pm 1$  models. We
address these and show that the big-bang singularity is resolved in
the same precise sense as in the recent analysis of the Bianchi I
model. Bianchi II space-times are of special interest to quantum
cosmology because of the expected behavior of the gravitational
field near \emph{generic} space-like singularities in classical
general relativity.

\end{abstract}


\maketitle

\section{Introduction}
\label{s1}

In this paper, we will study the loop quantum cosmology (LQC)
\cite{aa-rev,mbrev} of the Bianchi type II model. These models are
of special interest to the issue of singularity resolution
because of the intuition derived from the body of results related to
the Belinksii, Khalatnikov, Lifshitz (BKL) conjecture
\cite{bkl1,bkl2} on the nature of generic, spacelike singularities
in general relativity (see, e.g., \cite{bb}). Specifically, as the
system enters the Planck regime, dynamics at any fixed spatial point
is expected to be well described by the Bianchi I evolution.
However, there are transitions in which the parameters
characterizing the specific Bianchi I space-time change and the
dynamics of these transitions mimics the Bianchi II time evolution.
In a recent paper \cite{awe2}, we studied the Bianchi I model in the
context of LQC. In this paper we will extend that analysis to the
Bianchi II model. We will follow the same general approach and use
the same notation, emphasizing only those points at which the
present analysis differs from that of \cite{awe2}.

Bianchi I and II models are special cases of type A Bianchi models
which were analyzed already in the early days of LQC (see in
particular \cite{mb-hom, bdv}). However, as is often the case with
pioneering early works, these papers overlooked some important
conceptual and technical issues. At the classical level,
difficulties faced by the Hamiltonian (and Lagrangian) frameworks in
non-compact, homogeneous space-times went unnoticed. In these cases,
to avoid infinities, it is necessary to introduce an elementary cell
and restrict all integrals to it \cite{as,abl}. The Hamiltonian
frameworks in the early works did not carry out this step. Rather,
they were constructed simply by dropping an infinite volume integral
(a procedure that introduces subtle inconsistencies). In the quantum
theory, the kinematical quantum states were assumed to be periodic
---rather than almost-periodic--- in the connection, and the quantum
Hamiltonian constraint was constructed using a ``pre-$\mu_o$''
scheme. Developments over the intervening years have shown that
these strategies have severe limitations (see, e.g.,
\cite{aps3,acs,cs1,aa-badhonef,cs2}). In this paper, they will be
overcome using ideas and techniques that have been introduced in the
isotropic and Bianchi I models in these intervening years. Thus, as
in \cite{awe2} the classical Hamiltonian framework will be based on
a fiducial cell, quantum kinematics will be constructed using almost
periodic functions of connections and quantum dynamics will use the
``$\bar\mu$ scheme.'' Nonetheless, the space-time description of
Bianchi II models in \cite{mb-hom, bdv}, tailored to LQC, will
provide the point of departure of our analysis.

New elements required in this extension from the Bianchi I model
can be summarized as follows. Recall first that the spatially
homogeneous slices $M$ in Bianchi models are isomorphic to
3-dimensional group manifolds. The Bianchi I group is the
3-dimensional group of translations. Hence the the three Killing
vectors $\xiz^a_i$ on $M$ ---the left invariant vector fields on
the group manifold--- commute and coincide with the right
invariant vector fields $\e^a_i$ which constitute the fiducial
orthonormal triads on $M$. In LQC one mimics the strategy used in
LQG and spin foams and defines the curvature operator in terms of
holonomies around plaquettes whose edges are tangential to these
vector fields. The Bianchi II group, on the other hand, is
generated by the two translations and the rotation on a null
2-plane. Now the Killing vectors $\xiz^a_i$ no longer commute and
neither do the fiducial triads $\e^a_i$. Therefore we have to
follow another strategy to build the elementary plaquettes.
However, this situation was already encountered in the
k=$1$, isotropic models \cite{warsaw,apsv}. There, the desired
plaquettes can be obtained by alternating between the integral
curves of right and left invariant vector fields which do commute.
However, in the isotropic case, the gravitational connection is
given by $A_a^i = c\,\, \w_a^i$, where $\w_a^i$ are the covectors
dual to $\e^a_i$ and the holonomies around these plaquettes turned
out to be almost periodic functions of the connection component
$c$ \cite{warsaw,apsv}. By contrast, in the Bianchi II model we
have three connection components $c^i$ because of the presence of
anisotropies, and, unfortunately, the holonomies around our
plaquettes are no longer almost periodic functions of $c^i$. (This
is also the case in more complicated Bianchi models.) Since the
standard kinematical Hilbert space of LQC consists of almost
periodic functions of $c^i$, these holonomy operators are not
well-defined on this Hilbert space. Thus, the strategy \cite{abl}
used so far in LQC to define the curvature operator is no longer
viable.

One could simply enlarge the kinematical Hilbert space to
accommodate the new holonomy functions of connections. But then the
problem quickly becomes as complicated as full LQG. To solve the
problem within the standard, symmetry reduced kinematical framework
of LQC, one needs to generalize the strategy to define the curvature
operator. Of course, the generalization must be such that, when
applied to all previous models, it is compatible with the procedure
of computing holonomies around suitable plaquettes used there. We
will carry out this task by suitably modifying ideas that have
already appeared in the literature. This generalization will enable
one to incorporate \emph{all} class A Bianchi models in the LQC
framework.

Once this step is taken, one can readily construct the quantum
Hamiltonian constraint and the physical Hilbert space, following
steps that were introduced in the analysis \cite{awe2} of the
Bianchi I model. However, because Bianchi II space-times have
spatial curvature, the spin connection compatible with the
orthonormal triad is now non-trivial. It leads to two new terms in
the Hamiltonian constraint that did not appear in the Bianchi I
Hamiltonian. We will analyze these new terms in some detail. In
spite of these differences, the big bang singularity is resolved in
the same precise sense as in the Bianchi I model \cite{awe2}: If a
quantum state is initially supported only on classically
non-singular configurations, it continues to be supported on
non-singular configurations throughout its evolution.

The paper is organized as follows. Section \ref{s2} summarizes the
classical Hamiltonian theory describing Bianchi II models. Section
\ref{s3} discusses the quantum theory. We first define a non-local
connection operator $\hat{A}_a^i$ and use it to obtain the
Hamiltonian constraint. We then show that the singularity is
resolved and the Bianchi I quantum dynamics is recovered in the
appropriate limit. In Section \ref{s4}, we introduce effective
equations for the model (with the same caveats as in the Bianchi I
case \cite{awe2}).
Finally, in section V we summarize our results and discuss the new
elements that appear in the Bianchi II model. In Appendix A we
improve on the discussion of discrete symmetries presented in
\cite{awe2}. The results on the Bianchi I model obtained in
\cite{awe2} carry over without any change. But the change of
viewpoint is important to the LQC treatment of the Bianchi II model
and more general situations.

\section{Classical Theory}
\label{s2}

This section is divided into two parts. In the first we recall the
structure of Bianchi II space-times and in the second we summarize
the phase space formulation, adapted to LQC.

\subsection{Diagonal Bianchi II Space-times}
\label{s2.1}

Because the issue of discrete symmetries is subtle in background
independent contexts, and because it plays a conceptually important
role in the quantum theory of Bianchi II models, we will begin with
a brief summary of how various fields are defined
\cite{alrev,aa-dis}. This stream-lined discussion brings out the
assumptions which are often only implicit, making the discussion of
discrete symmetries clearer.

In the Hamiltonian framework underlying loop quantum gravity (LQG),
one fixes an \emph{oriented} 3-manifold $M$ and a 3-dimensional
`internal' vector space $I$ equipped with a positive definite metric
$q_{ij}$. The internal indices $i,j,k,\ldots$ are then freely
lowered and raised by $q_{ij}$ and its inverse. A spatial triad
$e^a_i$ is an isomorphism from $I$ to tangent space at each point of
$M$ which associates a vector field $v^a:= e^a_i v^i$ on $M$ to each
vector $v^i$ in $I$.%
\footnote{Thus, in LQG one begins with non-degenerate triads and
metrics, passes to the Hamiltonian framework and then, at the end,
extends the framework to allow degenerate geometries.}
The dual co-triads are denoted by $\omega_a^i$. Given a triad, we
acquire a positive definite metric $q_{ab}:= q_{ij} \omega_a^i
\omega_b^j$ on $M$. The metric $q_{ab}$ in turn singles out a 3-form
$\epsilon_{abc}$ on $M$ which has \emph{positive orientation} and
satisfies $ \epsilon_{abc}\epsilon_{def}\, q^{ad} q^{be}q^{cf}= 3!$.
One can then define a 3-form $\epsilon_{ijk}$ on $I$ via
$\epsilon_{ijk} = \epsilon_{abc} e^a_i e^b_j e^c_k$. Note that
$\epsilon_{ijk}$ is automatically compatible with $q_{ij}$, i.e.,
$\epsilon_{ijk}\epsilon_{lmn}\, q^{il} q^{jm} q^{kn}= 3!$. If a
triad $\bar{e}^a_i$ is obtained by flipping an odd number of the
vectors in the triad $e^a_i$, then $\bar{e}^a_i$ and $e^a_i$ have
opposite orientations and the fields they define satisfy
$\bar{q}_{ab} = {q}_{ab},\, \bar\epsilon_{abc} = \epsilon_{abc}$ but
$\bar\epsilon_{ijk} = - \epsilon_{ijk}$. Had we fixed
$\epsilon_{ijk}$ once and for all on $I$, then $\epsilon_{abc}$
would have flipped sign under this operation and volume integrals on
$M$ computed with the unbarred and barred triads would have had
opposite signs. With our conventions, these volume integrals will
not change and the parity flips will be symmetries of the symplectic
structure and the Hamiltonian constraint.

The triad also determines an unique spin connection $\Gamma_a^i$ via
\be \label{sc} D_{[a} \omega_{b]}^i\, \equiv \,
\partial_{[a}\omega_{b]}^i + \epsilon^{i}{}_{jk} \Gamma_{[a}^j
\omega_{b]}^k \, =\, 0\, .\ee
The gravitational configuration variable $A_a^i$ is then given by
$A_a^i = \Gamma_a^i + \gamma K_a^i$ where $K_{ab} := K_a^i
\omega_{bi}$ is the extrinsic curvature of $M$ and $\gamma$ is the
Barbero-Immirzi parameter, representing a quantization ambiguity.
(The numerical value of $\gamma$ is fixed by the black hole entropy
calculation.) The momenta $E^a_i$ carry, as usual, density weight 1
and are given by: $E^a_i = \sqrt{q} e^a_i$. The fundamental Poisson
bracket is:
\be \{A_a^i(x), \, E^b_j(y)\} = 8\pi G\gamma\,\, \delta_a^b\,
\delta^i_j\, \delta^3(x,y)\, .\ee

In Bianchi models \cite{taub,bianchi,atu}, one restricts oneself to
those phase space variables admitting a 3-dimensional group of
symmetries which act simply and transitively on $M$. Thus, the
3-metrics $q_{ab}$ under consideration admit a 3-parameter group of
isometries and $M$ is diffeomorphic to a 3-dimensional Lie group
$G$. (However, there is no canonical diffeomorphism, so that there
is no preferred point on $M$ corresponding to the identity element
of $G$.) To avoid a proliferation of spaces and types of indices, it
is convenient to identify the internal space $I$ and the Lie-algebra
$\L G$ of $G$ via a fixed isomorphism. Then, there is a natural
isomorphism $\xiz^a_i$ between $\L G \equiv I$ and Killing vector
fields on $M$: for each internal vector $v^i$, $\xiz^a_i v^i$ is a
Killing field on $M$. For brevity we will refer to $\xiz^a_i$ as
(left invariant) vector fields on $M$. There is a canonical triad
$\e^a_i$
---the right invariant vector fields--- which is Lie dragged by the
$\xiz^a_i$. This triad and the dual co-triad $\w_a^i$ satisfy:
\ba [ \xiz_i,\, \e_j ] &=&0, \quad\quad  [\e_i,\, \e_j] =
- \C_{ij}^k\, \e_k,\nonumber\\
\L_{\xiz_i}\,( \w^j) &=&0, \quad\quad \dd\,\w^k = \f{1}{2}\,
\C_{ij}^k \w^i\wedge\w^j,\ea
where $\C_{ij}^k$ denotes the structure constants of $\L G$. It is
convenient to use the fixed fields $\e^a_i$ and $\w_a^i$ as
\emph{fiducial} triads and co-triads.

In the case when $G$ is the Bianchi II group, we have $\C_{ik}^k =0$
as in all class A Bianchi models and, furthermore, the symmetric
tensor $k^{kl}:=\C_{ij}^k \, \epsilon^{ijl}$ has signature +,0,0.
Therefore, we can fix, once and for all an orthonormal basis
$\b_1^i, \b_2^i, \b_3^i$ in $I$ such that the only non-zero
components of $\C_{ij}^k$ are
\be \C_{23}^1 = - \C_{32}^1 = \ta\, ,\ee
where $\ta$ is a non-zero real number.%
\footnote{Without loss of generality $\ta$ can be chosen to be 1. We
keep it general because we will rescale it later (see Eq.
(\ref{tilde})) and because we want to be able to pass to the Bianchi
I case by taking the limit $\ta\to0$.}
We will assume that this basis is so oriented that
\be \label{ve1} \epsilon_{123}\, :=\, \epsilon_{ijk} \, \b^i_1\,
\b^j_2\, \b^k_3\, \, =\, \ve\ee
where $\ve = \pm 1$ depending on whether the frame $e^a_i$ (which
determines the sign of $\epsilon_{ijk}$) is right or left handed.
Throughout this paper we will set $\xiz^a_1 = \xiz^a_i \b^i_1,\,
\e^a_1 = \e^a_i\b^i_1,\, \w_a^1 = \w_a^i\b^1_i$, etc.

The form of the components of $\C^k_{ij}$ in this basis implies that
$M$ admits global coordinates $x,y,z$ such that the Bianchi II
Killing vectors have the fixed form
\be \xiz_1^a = \left(\f{\partial}{\partial x}\right)^a, \qquad
 \xiz^a_2 = \left(\f{\partial}{\partial y}\right)^a, \qquad
 \xiz^a_3 = \ta y\left(\f{\partial} {\partial x}\right)^a+
 \left(\f{\partial}{\partial z}\right)^a. \ee
These expressions bring out the fact that, if we were to attempt to
compactify the spatial slices to pass to a $\mathbb{T}^3$ topology
---as one can in the Bianchi I model--- we will no longer have
globally well-defined Killing fields. Thus, in the Bianchi II model,
we are forced to deal with the subtleties associated with
non-compactness of the spatially homogeneous slices.

In the $x,y,z$ chart, the right invariant triad is given by
\be \e^a_1 = \left(\f{\partial}{\partial x}\right)^a, \qquad \e^a_2
= \ta z \left(\f{\partial}{\partial
x}\right)^a+\left(\f{\partial}{\partial y} \right)^a, \qquad \e^a_3
= \left(\f{\partial}{\partial z}\right)^a, \ee
and the dual co-triad by
\be \w_a^1=(\dd x)_a-\ta z(\dd y)_a, \qquad\w_a^2=(\dd y)_a,
\qquad\w_a^3=(dz)_a. \ee
They determine a fiducial 3-metric $\q_{ab}:= q_{ij}\w_a^i\w_b^j$
with Bianchi II symmetries:
\be \q_{ab} \dd x^a \dd x^b =  (\dd x-\ta z\:\dd y)^2\,+\,\dd
y^2\,+\, \dd z^2. \ee

In the diagonal models, the physical triads $e^a_i$ are related to
the fiducial ones by%
\footnote{There is no sum if repeated indices are both covariant or
contravariant.  As usual, the Einstein summation convention holds if a
covariant index is contracted with a contravariant index.}
\be \label{edef} \omega_a^i = a^i(\tau)\w_a^i, \qquad \mathrm{and}
\qquad a_i(\tau) e^a_i = \e^a_i
 \ee
where the $a_i$ are the three directional scale factors. Since the
physical spatial metric is given by  $q_{ab} =
\omega_a^i\omega^{}_{bi}$, the space-time metric can be expressed as
\be \label{metric} \dd s^2= -N \dd\tau^2 + a_1(\tau)^2\:(\dd x-\ta
z\:\dd y)^2+a_2(\tau)^2\:\dd y^2+a_3(\tau)^2\:\dd z^2 \ee
where $N$ is the lapse function adapted to the time coordinate
$\tau$.

For later use, let us calculate the spin connection (\ref{sc})
determined by triads $e^a_i$. From the definition of $\Gamma_a^i$ it
follows that
\be \Gamma_a^i =
-\epsilon^{ijk}\,e^b_j\,\left(\partial_{[a}\omega_{b]k}+\f{1}{2}e^c_k
\omega^l_a\partial_{[c}\omega_{b]l}\right)\, . \ee
Using (\ref{ve1}), the components of $\Gamma_a^i$ in the internal
basis $\b^i_1, \b^i_2, \b^i_3$ can be expressed as
\be \Gamma_a^1 = \f{\ta\ve a_1^2}{2a_2a_3}\:\w_a^1; \qquad \Gamma_a^2
=-\f{\ta\ve a_1}{2a_3}\:\w_a^2; \qquad \Gamma_a^3 = -\f{\ta\ve
a_1}{2a_2}\:\w_a^3. \ee

Before studying the dynamics of the model, let us examine the action
of internal parity transformation $\Pi_k$ which flips the $k$th
triad vector and leaves the orthogonal vectors alone. (For details
see Appendix and \cite{aa-dis}). Under the parity transformation
$\Pi_1$, for example, we have: $e^a_1\,\to \, -e^a_1,\, e^a_2\, \to
e^a_2,\, e^a_ 3\, \to \, e^a_3$ and $a_1\to -a_1,\, a_2\to a_2, a_3
\to a_3$ whence $\Gamma_a^1 \to -\Gamma^a_1,\, \Gamma_a^2 \to
\Gamma^a_2,\, \Gamma_a^3 \to \Gamma^a_3$. Thus, both $e^a_i$ and
$\Gamma_a^i$ are \emph{proper} internal vectors. $\ve$ on the other
hand is a pseudo internal scalar, $\ve \to -\ve$ under every
$\Pi_k$. Note that the fiducial quantities carrying a label $o$ do
not change under this transformation; it affects only the physical
quantities.

\subsection{The Bianchi II Phase space}
\label{s2.2}

As is usual in LQC, we will now use the fiducial triads and
co-triads to introduce a convenient parametrization of the phase
space variables, $E^a_i, A_a^i$. Because we have restricted
ourselves to the diagonal model and these fields are symmetric under
the Bianchi II group, from each equivalence class of gauge related
phase space variables we can choose a pair of the form
\be \label{var} E^a_i = \tp_i\sqrt{|\q|}\,\e^a_i \qquad \mathrm{and}
\qquad A_a^i = \tc^i \,\w_a^i, \ee
where, as spelled out in footnote 3, there is no sum over $i$. Thus,
a point in the phase space is now coordinatized by six real numbers
$\tp_i,\tc^i$. One would now like to use the symplectic structure in
full general relativity to induce a symplectic structure on our
six-dimensional phase space. However, because of spatial homogeneity
and the ${\R}^3$ spatial topology, the integrals defining the
symplectic structure, the Hamiltonian (and the action) all diverge.
Therefore we have to introduce a fiducial cell $\mathcal{V}$ and
restrict integrals to it \cite{as,abl}. We will take the fiducial
cell to be rectangular with edges along the coordinate axes and
lengths of $L_1, L_2$ and $L_3$ with respect to the \emph{fiducial}
metric $\q_{ab}$. It then follows that the volume of the fiducial
cell with respect to $\q_{ab}$ is $V_o=L_1L_2L_3$. Then the non-zero
Poisson brackets are given by:
\be \label{pb1} \{\tc^i,\, \tp_j\} \, = \, \f{8\pi G \gamma}{V_o}\,
\delta^i_j \ee
where $\gamma$ is the Barbero-Immirzi parameter. As in the Bianchi I
case, we have a 1-parameter ambiguity in the symplectic structure
because of the explicit dependence on $V_o$ and we have to make sure
that the final physical results are either independent of $V_o$ or
remain well-defined as we remove the `regulator' and take the limit
$V_o \to \infty$.

It is convenient to rescale variables to absorb this dependence in
the phase space coordinates (as was done in the treatment of Bianchi
I model in \cite{awe2}). Let us set
\be p_1= L_2L_3\tp_1, \qquad p_2=L_3L_1\tp_2, \qquad p_3= L_1L_2\tp_3,
\ee
\be \label{tilde} c_1=L_1\tc_1, \qquad c_2=L_2\tc_2, \qquad
c_3=L_3\tc_3 \qquad \mathrm{and} \qquad \alpha =
\f{L_2L_3}{L_1}\ta\, , \ee
where the last rescaling has been introduced to absorb factors of
$L_i$ which would otherwise unnecessarily obscure the expression of
the Hamiltonian constraint. The Poisson brackets between these new
phase space coordinates is given by%
\be \label{pb2}\{c^i,\, p_j\} \, = \, 8\pi G \gamma \,\delta^i_j \,
. \ee
These variables have direct physical interpretation. For example,
$p_1$ is the (oriented) area of the 2-3 face of the elementary cell
with respect to the \emph{physical} metric $q_{ab}$ and $h^{(1)} =
\exp c_1\tau_1$ is the holonomy of the physical connection $A_a^i$
along the first edge of the elementary cell.

Our choice (\ref{var}) of physical triads and connections has fixed
the internal gauge as well as the diffeomorphism freedom.
Furthermore, it is easy to explicitly verify that, thanks to
(\ref{var}), the Gauss and the diffeomorphism constraints are
automatically satisfied. Thus, as in \cite{awe2}, we are left just
with the Hamiltonian constraint
\be \label{Hgen} \mathcal{C}_H = \int_\mathcal{V}
\Big[\f{NE^a_iE^b_j}{16\pi G\sqrt{|q|}}
\big(\epsilon^{ij}{}_kF_{ab}{}^k-2(1+\gamma^2)K_{[a}^iK_{b]}^j \Big)
+ N \mathcal{H}_{\m}\big]\, \dd^3x\, , \ee
where
\be F_{ab}{}^k=2\partial_{[a}A_{b]}^k+\epsilon_{ij}{}^kA_a^iA_b^j \ee
is the curvature of $A_a^i$ and $\mathcal{H}_\m$ is the matter
Hamiltonian density. As in \cite{awe2}, our matter field will
consist only of a massless scalar field $T$ which will later serve
as a relational time variable a la Liebniz. (Additional matter
fields can be incorporated in a straightforward manner, modulo
possible intricacies of essential self-adjointness.) Thus,
\be \mathcal{H}_{\m} = \f{1}{2}\f{p_T^2}{\sqrt{|q|}}. \ee

Since we want to use the massless scalar field as relational time,
it is convenient to use a harmonic-time gauge, i.e., assume that the
time coordinate $\tau$ in (\ref{metric}) satisfies $\Box \tau=0$.
The corresponding lapse function is $N=\sqrt{|p_1p_2p_3|}$. With
this choice, the Hamiltonian constraint simplifies considerably.
Note first that the basic canonical variables can be expanded as
\be E^a_i = \f{p_i}{V_o}L_i\sqrt{|\q|}\e^a_i \qquad {\rm and} \qquad
A_a^i = \f{c^i}{L^i}\w_a^i, \ee
and the extrinsic curvature is given by
 $$\qquad  K_a^i = \gamma^{-1} (A_a^i-\Gamma_a^i).$$
Next, using $p_1 = ({\rm sgn}a_1)\, |a_2a_3|\,L_2L_3$ etc, the
components of the spin connection become:
\be \Gamma_a^1 = \f{\alpha\ve p_2p_3}{2p_1^2}\f{\w_a^1}{L_1}, \qquad
\Gamma_a^2= -\f{\alpha\ve p_3}{2p_1}\f{\w_a^2}{L_2}, \qquad
\Gamma_a^3=-\f{\alpha\ve p_2}{2p_1} \f{\w_a^3}{L_3} \, .\ee
Collecting terms, the Hamiltonian constraint (\ref{Hgen}) becomes
\begin{align} \label{Hcl} \mathcal{C}_H&=-\f{1}{8\pi G\gamma^2}
\Big[p_1p_2c_1c_2+p_2p_3 c_2c_3+p_3p_1c_3c_1+\alpha\ve p_2p_3c_1
\nonumber \\
&\qquad\qquad-(1+ \gamma^2)\,\big(\f{\alpha p_2p_3}{2p_1}\big)^2\Big]
 + \f{1}{2}p_T^2 \\
& = \mathcal{C}_H^{\rm (BI)} - \f{1}{8\pi G\gamma^2}\Big[\alpha\ve
p_2p_3c_1-(1+ \gamma^2)\,\big(\f{\alpha p_2p_3}{2p_1}\big)^2\Big],
\end{align}
where $\mathcal{C}_H^{\rm (BI)}$ is the Hamiltonian constraint
(including the matter term) for Bianchi I space-times which has
already been studied in \cite{awe2}.  Note that this constraint is
recovered in the limit $\alpha\to 0$, as it must be.

Knowing the form of the Hamiltonian constraint, it is now possible to derive
the time evolution of any classical observable $\mathcal{O}$ by taking its
Poisson bracket with $\mathcal{C}_H$:
\be \dot{\mathcal{O}} = \{\mathcal{O},\mathcal{C}_H\}\, , \ee
where the `dot' stands for derivative with respect to harmonic time
$\tau$. This gives
\be \label{ceom1} \dot{p_1}=\gamma^{-1}(p_1p_2c_2+p_1p_3c_3+\alpha\ve p_2p_3), \ee
\be \dot{p_2}=\gamma^{-1}(p_2p_1c_1+p_2p_3c_3), \ee
\be \dot{p_3}=\gamma^{-1}(p_3p_1c_1+p_3p_2c_2), \ee
\be
\dot{c_1}=-\f{1}{\gamma}\Big(p_2c_1c_2+p_3c_1c_3+\f{1}{2p_1}(1+\gamma^2)
\big(\f{\alpha p_2p_3}{p_1}\big)^2\Big), \ee
\be \dot{c_2}=-\f{1}{\gamma}\Big(p_1c_2c_1+p_3c_2c_3+\alpha\ve
p_3c_1-\f{1}{2p_2} (1+\gamma^2)\big(\f{\alpha
p_2p_3}{p_1}\big)^2\Big), \ee
\be \label{ceom2}
\dot{c_3}=-\f{1}{\gamma}\Big(p_1c_3c_1+p_2c_3c_2+\alpha\ve p_2
c_1-\f{1}{2p_3}(1+\gamma^2)\big(\f{\alpha p_2p_3}{p_1}\big)^2\Big).
\ee
Any initial data satisfying the Hamiltonian constraint can be
evolved by using the six equations above. It is straightforward to
extend these results if there are additional matter fields.

Finally, let us consider the parity transformation $\Pi_k$ which
flips the $k$th \emph{physical} triad vector $e^a_k$. (As noted
before, this transformation does not act on any of the fiducial
quantities which carry a label $o$.) Under this map, we have:
$q_{ab} \to q_{ab}, \, \epsilon_{abc} \to \epsilon_{abc}\,$ but
$\epsilon_{ijk} \to -\epsilon_{ijk}, \, \ve \to -\ve$. The canonical
variables $c^i, p_i$ transform as proper internal vectors and
co-vectors: For example
\be \Pi_1(c_1,c_2,c_3) \rightarrow (-c_1, c_2, c_3) \qquad {\rm and}
\qquad \Pi_1(p_1,p_2,p_3) \rightarrow (-p_1, p_2, p_3)\, . \ee
Consequently, both the symplectic structure and the Hamiltonian
constraint are left invariant under any of the parity maps $\Pi_k$.

This Hamiltonian description will serve as the point of departure
for loop quantization in the next section.

\section{Quantum Theory}
\label{s3}

This section is divided into three parts.  In the first, we discuss
the kinematics of the model, in the second we define an operator
corresponding to the connection $A_a^i$ using holonomies and in the
third we introduce the Hamiltonian constraint operator and describe
its action on states.

\subsection{LQC Kinematics}

The kinematics for the LQC of Bianchi II models is almost identical
to that for Bianchi I models. Therefore, in the sub-section we
closely follow \cite{awe2}.

Let us begin by specifying the elementary functions on the
classical phase space which will have unambiguous analogs in the
quantum theory. As in the Bianchi I model, the elementary
variables are the momenta $p_i$ and holonomies of the
gravitational connection $A_a^i$ along the integral curves of the
right invariant vector fields $\e^a_i$. Let $\tau_i$ be a basis of
the Lie algebra of SU(2), satisfying $\tau_i \tau_j =
\f{1}{2}\epsilon_{ij}{}^k \tau_k- \f{1}{4} \delta_{ij}\mathbb{I}$
where $\mathbb{I}$ is the unit $2\times2$ matrix. Consider an edge
of length $\ell L_k$ with respect to the fiducial metric
$\q_{ab}$, parallel to $\e^a_k$. The holonomy $h_k^{(\ell)}$ along
it is given by
\be \label{hol} h_k^{(\ell)}(c_1,c_2,c_3) = \exp\left(\ell
c_k\tau_k\right) = \cos\f{\ell c_k}{2} \mathbb{I} + 2\sin\f{\ell
c_k}{2}\tau_k. \ee
(Note that $\ell$ depends of the fiducial cell but not on the
fiducial metric.) This family of holonomies is completely
determined by the almost periodic functions $\exp(i\ell c_k)$ of
the connection. These almost periodic functions will be our
elementary configuration variables which will be promoted
unambiguously to operators in the quantum theory.

It is simplest to use the $p$-representation to  specify the
gravitational sector $\Hkg$ of the kinematic Hilbert space. The
orthonormal basis states $|p_1,p_2,p_3\rangle$ are eigenstates of
quantum geometry. For example, in the state $|p_1,p_2,p_3\rangle$
the face $S_{23}$  of the fiducial cell $\mathcal{V}$ (given by $x$
={\rm const}) has area $|p_1|$.
The basis is orthonormal in the sense
\be \langle p_1,p_2,p_3|p_1',p_2',p_3'\rangle = \delta_{p_1^{}p_1'}
\delta_{p_2^{}p_2'}\delta_{p_3^{}p_3'}\, , \ee
where the right side features Kronecker symbols rather than the
Dirac delta distributions. Hence kinematical states can consist only
of \emph{countable} linear combinations
\be |\Psi\rangle \,=\,
\sum_{p_1,p_2,p_3}\Psi(p_1,p_2,p_3)|p_1,p_2,p_3\rangle\ \ee
of these basis states for which the norm
\be \label{norm} ||\Psi ||^2\, =\, \sum_{p_1,p_2,p_3}\,
|\Psi(p_1,p_2,p_3)|^2 \ee
is finite. Because the right side features a sum over a countable
number of points on ${\R}^3$, rather than a 3-dimensional integral,
LQC kinematics are inequivalent to those of the Schr\"odinger
approach used in Wheeler-DeWitt quantum cosmology.

Next, recall that on the classical phase space the three reflections
$\Pi_i:\,\,e^a_i\,\to\, -e^a_i$ are large gauge transformations
under which physics does not change (since both the metric and the
extrinsic curvature are left invariant). These large gauge
transformations have a natural induced action, denoted by
$\hat\Pi_i$, on the space of wave functions $\Psi(p_1,p_2,p_3)$. For
example,
\be \hat\Pi_1\Psi(p_1,p_2,p_3)=\Psi(-p_1,p_2,p_3). \ee
Since $\hat\Pi_i^2$ is the identity, for each $i$, the group of
these large gauge transformations is simply $\Z_2$. As in Yang-Mills
theory, physical states belong to its irreducible representation.
For definiteness, as in the isotropic and Bianchi I models, we will
work with the symmetric representation. It then follows that
$\mathcal{H}_{\mathrm{kin}}^{\mathrm{grav}}$ is spanned by wave
functions $\Psi(p_1,p_2,p_3)$ which satisfy
\be \label{parity} \Psi(p_1,p_2,p_3)=\Psi(|p_1|,|p_2|,|p_3|) \ee
and have a finite norm (\ref{norm}).

The action of the elementary operators on
$\mathcal{H}_{\mathrm{kin}}^{\mathrm{grav}}$ is as follows: the
momenta act by multiplication whereas the almost periodic
functions in $c_i$ shift the $i$th argument. For example,
\be [\hat p_1 \Psi](p_1,p_2,p_3) = p_1\, \Psi(p_1,p_2,p_3) \,\quad
\mathrm{and} \,\quad \Big[\widehat{\exp(i\ell c_1)}\Psi\Big](p_1,
p_2, p_3) = \Psi(p_1-8\pi\gamma G\hbar \ell, p_2, p_3)\, . \ee
The expressions for $\hat p_2, \widehat{\exp(i\ell c_2)}, \hat
p_3$ and $\widehat{\exp(i\ell c_3)}$ are analogous.  Finally, we
need to define the operator $\hat{\ve}$ since $\ve$ features in
the expression of the Hamiltonian constraint. In the classical
theory, $\ve$ is unambiguously defined on non-degenerate triads,
i.e., when $p_1p_2p_3 \not= 0$. In quantum theory, wave functions
can have support also on degenerate configurations. We will extend
the definition to degenerate triads using the basis
$|p_1,p_2,p_3\rangle$:
\be \label{ve2} \hat{\ve}\,|p_1,p_2,p_3\rangle := \left\{
\rlap{\raise2ex\hbox{\,\,$\quad|p_1,p_2,p_3 \rangle$ if $p_1p_2p_3
\ge 0$,}}{\lower2ex\hbox{\,\,$ -\,|p_1,p_2,p_3 \rangle$ if
$p_1p_2p_3<0$.}} \right. \ee
Finally, the full kinematical Hilbert space
$\mathcal{H}_{\mathrm{kin}}$ will be the tensor product
$\mathcal{H}_{\mathrm{kin}}=\mathcal{H}_{\mathrm{kin}}^
{\mathrm{grav}}\otimes\mathcal{H}_{\mathrm{kin}}^{\mathrm{matt}}$,
where $\mathcal{H}_{\mathrm{kin}}^{\mathrm{matt}}=L^2({\R},dT)$ is
the matter kinematical Hilbert space for the homogeneous scalar
field.  On $\mathcal{H}_ {\mathrm{kin}}^{\mathrm{matt}}$, $\hat T$
will act by multiplication and $\hat p_T:=-i\hbar \mathrm{d}_T$
will act by differentiation. As in isotropic and Bianchi I models,
our final results would remain unaffected if we use a ``polymer
representation'' also for the scalar field.

\subsection{The connection operator $\hat{A}_a^i$}
\label{s3.2}

To define the quantum Hamiltonian constraint, we cannot directly use
the symmetry reduced classical constraint (\ref{Hcl}) because it
contains connection components $c_k$ themselves and in LQC only
almost periodic functions of $c_k$ have unambiguous operator
analogs. Indeed, in all LQC models considered so far
\cite{abl,aps3,warsaw,apsv,kv1,ls,bp,awe2}, we were led to return to
the expression (\ref{Hgen}) in the full theory and mimic the
procedure used in LQG \cite{tt}. More precisely, the key strategy
was to follow full LQG (and spin foams) and define a ``field
strength operator'' using holonomies around suitable closed loops.
In the Bianchi I model, these closed loops were formed by following
integral curves of right invariant vector fields (which are also
left invariant). As mentioned in section \ref{s2}, in the Bianchi II
model the right invariant vector fields define the fiducial triads
$\e^a_i$, the left invariant vector fields, the Killing fields
$\xiz^i$. Neither constitutes a commuting set, whence their integral
curves cannot be used to form closed loops. However, as in the k=1
case \cite{warsaw,apsv}, one can hope to exploit the fact that the
right invariant vector fields do commute with the left invariant
ones and construct the closed loops by alternately following right
and left invariant vector fields. But, as mentioned in section
\ref{s1}, a new problem arises: unlike in the k=1 (or Bianchi I)
model the resulting holonomies are no longer almost periodic
functions of $c_k$, whence the Hilbert space $\H_{\rm kin}^{\rm
grav}$ does not support these holonomy operators. For completeness
we will first show this fact explicitly and then introduce a new
avenue to bypass this difficulty.

The problematic curvature component turns out to be $F_{yz}{}^1$. To
construct the corresponding operator, following the strategy used in
the k=1 case \cite{warsaw,apsv}, we will construct a closed loop
$\Box_{yz}$ as follows. In the coordinates $(x,y,z)$,\,\, i) Move
from $(0,0,0)$ to $(0,\bar\mu_2L_2,0)$ following $\xi^a_2$;\,\, ii)
then move from $(0,\bar\mu_2L_2,0)$ to
$(0,\bar\mu_2L_2,\bar\mu_3L_3)$ following $\e^a_3$;\,\, iii) then
move from $(0,\bar\mu_2L_2,\bar\mu_3L_3)$ to $(0,0,\bar\mu_3L_3)$
following $-\xi^a_2$;\,\, and, finally, iv) move from
$(0,0,\bar\mu_3L_3)$ to $(0,0,0)$ following $-\e^a_3$. The
parameters $\bar\mu_i$ which determine the `lengths' of these edges
can be fixed by the semi-heuristic correspondence between LQC and
LQG exactly as in the Bianchi I model \cite{awe2} because the
geometric considerations used in that analysis continue to hold
without any modification in all Bianchi models with $\R^3$ spatial
topology:
\be \label{mubar} \bar\mu_1 =
\sqrt\f{|p_1|\Delta\,\lp^2}{|p_2p_3|}, \qquad  \bar\mu_2 =
\sqrt\f{|p_2|\Delta\,\lp^2}{|p_1p_3|}, \qquad \bar\mu_3 =
\sqrt\f{|p_3| \Delta\,\lp^2}{|p_1p_2|} \ee
where $\Delta\,\lp^2 = 4\sqrt{3}\pi\gamma\,\lp^2$ is the `area gap'.
The holonomy around this closed loop $\Box_{yz}$ is given by
\be {h}_{\Box_{yz}} = \f{2}{c\,\,\bar\mu_2\bar\mu_3L_2L_3}\cos\left(
\f{\bar\mu_2c_2}{2}\right)\sin\left(\f{\bar\mu_2 c}{2}\right)
\Big(c_2\sin(\bar\mu_3c_3)+
\alpha\bar\mu_3c_1\cos(\bar\mu_3c_3)\Big) \ee
where
\be \label{c12} c = \sqrt{\alpha^2\bar\mu_3^2c_1^2+c_2^2}. \ee
If we were to shrink the loop so that the area it encloses goes to
zero, we do indeed recover the classical expression of $F_{yz}{}^1$.
However, because of presence of the term $c$, if $\alpha\not=0$ the
right side fails to be almost periodic in $c_1$ and $c_2$. Hence
this holonomy operator fails to exist on $\H_{\rm kin}$. It is clear
from the expression (\ref{c12}) of $c$ that the problem is
independent of the specific way $\bar\mu_i$ are fixed.

We will bypass this difficulty by mimicking another strategy used in
full LQG \cite{tt}: We will use holonomies along segments parallel
to $\e^a_i$ to define an operator corresponding to the connection
itself. This is a natural strategy because holonomies along these
segments suffice to separate the Bianchi II connections (\ref{var}).
Let us set $A_a := A_a^k\tau_k$. Then we have the identity:
\be \label{classA} A_a =  \lim_{\ell_k \to 0}\, \sum_k
\,\f{1}{2\ell_kL_k}\,\, \Big(h_k^{(\ell_k)} -
(h_k^{(\ell_k)})^{-1}\Big) \ee
where $h_k^{(\ell_k)}$ is given by (\ref{hol}). In the expressions
of physically interesting operators such as the Hamiltonian
constraint of full LQG, one often replaces $A_a$ with the (analog of
the) right side of (\ref{classA}). But because of the specific forms
of these operators, the limit trivializes on diffeomorphism
invariant states of LQG. In LQC, we have gauge fixed the system and
therefore cannot appeal to diffeomorphism invariance. Indeed, while
the holonomies are well-defined for each non-zero $\ell_k$, the
limit fails to exist on $\H_{\rm kin}^{\rm grav}$. A natural
strategy is to shrink $\ell_k$ to a judiciously chosen non-zero
value. But what would this value be? In the case of plaquettes, we
could use the interplay between LQG and LQC directly because the
argument $p_i$ of LQC quantum states refers to \emph{quantum} areas
of faces of the elementary cell $\mathcal{V}$ \cite{awe2}. For edges
we do not have such direct guidance. There is, nonetheless a natural
principle one can adopt: Normalize $\ell_k$ such that the numerical
coefficient in front of the curvature operator constructed from the
resulting connection agrees with that in the expression of the
curvature operator constructed from holonomies around closed loops,
in all cases where the second construction is available. We will use
this strategy. Let us apply it to the Bianchi I model where
$F_{ab}{}^k = \epsilon_{ij}{}^k\, A_a^i A_b^j$. Using holonomies
around closed loops one obtains the field strength operator
\be \hat{F}_{ab}{}^k = \epsilon_{ij}{}^k\,
\big(\f{\sin\bar{\mu}c}{\bar{\mu}L}\, \w_a\big)^i\,
\big(\f{\sin\bar{\mu}c}{\bar{\mu}L}\, \w_b\big)^j \ee
where
\be \big(\f{\sin\bar{\mu}c}{\bar{\mu}L}\, \w_a\big)^i =
\big(\f{\sin\bar{\mu_i}c_i}{\bar{\mu_i}L_i}\, \w_a^i\big) \quad\quad
\hbox{\rm (no sum over i)} \nonumber \ee
(see Eqs (3.12) and (3.13) in \cite{awe2}). Therefore, our strategy
yields $\ell_k = 2\bar\mu_k$, that is,
\be \label{Aop} \hat{A}_a^k =
\f{\widehat{\sin(\bar\mu^kc^k)}}{\bar\mu^kL_k}\,\,\w_a^k, \ee
where there is no sum over $k$. Note that the principle stated above
leads us unambiguously to the factor $2$ in $\ell_k = 2\bar\mu_k$;
without recourse to a systematic strategy, one may have naively set
$\ell_k =\bar\mu_k$.

If we compare the expression (\ref{Aop}) of the connection operator
with the expression (\ref{var}) of the classical connection, we have
effectively defined an operator $\hat{c}$ via
\be \hat{c}_k = \f{\widehat{\sin(\bar\mu^kc^k)}}{\bar\mu^k} \ee
where there is again no sum over $k$. In the literature such a
quantization of $c$ is often called ``polymerization.'' Our approach
is an improvement over such strategies in two respects. First, we
did not just make the substitution $c \rightarrow \sin \ell c/\ell$
by hand; a priori one could have used another substitution such as
$c \rightarrow \tan \ell c/\ell$. Rather, as in full LQG, we used
the strategy of expressing the connection in term of holonomies,
`the elementary variables'. But this still leaves open the question
of what $\ell$ one should use. We determined this by requiring that
the overall normalization of $\hat{F}_{ab}{}^k$ constructed from
$\hat{A}_a^i = c^i (L^i)^{-1}\,\w_a^i$ should agree with that of
$\hat{F}_{ab}{}^k$ constructed from holonomies around appropriate
closed loops, when the second construction is possible. Therefore,
our construction is a bona-fide generalization of the previous
constructions used successfully in LQC.

This strategy has some applications beyond the Bianchi II model
studied in this paper. First, the k=$-1$ isotropic case has been
studied in detail in \cite{kv1,ls}. The analysis uses the $\bar\mu$
scheme, carries out a numerical simulation using exact LQC equations
and shows that the effective equations of the ``embedding approach"
\cite{jw,vt} (discussed in section \ref{s4}) provide an excellent
approximation to the quantum evolution. While this is an essentially
exhaustive treatment, as \cite{kv1,ls} itself points out, the
treatment has a conceptual limitation in that it builds holonomies
around the closed loops using the extrinsic curvature $K_a^i$
---rather than $A_a^i$--- as a ``connection''. This limitation can
be overcome in a straightforward fashion using our current strategy.
More importantly, this strategy is applicable to all class A Bianchi
models, including type IX. Thus, it opens the door to the LQC
treatment of all these models in one go.

\subsection{The quantum Hamiltonian constraint}
\label{s3.3}

With the connection operator at hand, one can construct the
Hamiltonian constraint operator starting either from the general LQG
expression (\ref{Hgen}) or the symmetry reduced expression
(\ref{Hcl}). We will begin by a small change in the representation
of kinematical states which will facilitate this task.

\subsubsection{A more convenient representation}
\label{s3.3.1}

Ignoring factor ordering ambiguities for the moment, the
constraint operator $\hat{\mathcal{C}}_H$ is given by
\begin{align} \label{qHam1} \hat{\mathcal{C}}_H = -\f{1}{8\pi
G\gamma^2\Delta\lp^2}&\Big[p_1 p_2|p_3|\sin\bar\mu_1c_1
\sin\bar\mu_2c_2+|p_1|p_2p_3\sin\bar\mu_2c_2\sin\bar\mu_3c_3 \nonumber \\
&+p_1|p_2|p_3\sin\bar\mu_3c_3\sin\bar\mu_1c_1\Big]-
\f{1}{8\pi G\gamma^2}\Big[\alpha\hat{\ve}p_2p_3\sqrt\f{|p_2p_3|}{|p_1|\Delta
\lp^2}\sin\bar\mu_1c_1\nonumber \\ & -(1+\gamma^2)\left(\f{\alpha
p_2p_3}{2p_1}\right)^2\Big]+\f{1}{2}\hat{p}_T^2 \end{align}
where for simplicity of notation here and in what follows we have
dropped the hats on the $p_i$ and $\sin\bar\mu_ic_i$ operators.
Recall that, classically, the Bianchi II symmetry group reduces to
the Bianchi I symmetry group if we set $\alpha=0$. If one sets
$\alpha=0$ in (\ref{qHam1}), the last two terms disappear and the
operator $\hat{\mathcal{C}}_H$ reduces to that of the Bianchi I
model \cite{awe2} showing explicitly that our construction is a
natural generalization of the strategy used there.

To obtain the action of operators corresponding to terms of the form
$\sin\bar\mu_ic_i$ we use the same strategy as in \cite{awe2}. As
shown there, it is simplest to introduce dimensionless variables
\be
\l_i=\f{\sgn(p_i)\sqrt{|p_i|}}{(4\pi\gamma\sqrt\Delta\lp^3)^{1/3}}\,
. \ee
%
%
Then the kets $|\l_1,\l_2,\l_3\rangle$ constitute an orthonormal
basis in which the operators $p_k$ are diagonal
\be p_k|\l_1,\l_2,\l_3\rangle\, =\,
[\sgn(\l_k)(4\pi\gamma\sqrt\Delta\lp^3)^{2/3}
\l_k^2]\,\,|\l_1,\l_2,\l_3\rangle\, . \ee
Quantum states will now be represented by functions
$\Psi(\l_1,\l_2,\l_3)$. The operator $e^{i\bar\mu_1c_1}$ acts on
them as follows
\begin{align} \big[e^{i\bar\mu_1c_1}\,\Psi\big] (\l_1,\l_2,\l_3)
&= \Psi(\l_1- \f{1}{|\l_2\l_3|},\l_2,\l_3) \nonumber \\
&= \Psi(\f{v-2\sgn(\l_2\l_3)}{v}\cdot\, \l_1,\l_2,\l_3),
\end{align}
where we have introduced the variable $v=2\l_1\l_2\l_3$ which is
proportional to the volume of the fiducial cell:
\be \hat{V}\,\Psi(\l_1,\l_2,\l_3)\, =\,
[2\pi\gamma\sqrt\Delta\,|v|\,\lp^3]\, \Psi(\l_1, \l_2,\l_3). \ee
(The $e^{i\bar\mu_1c_1}$ operator is well-defined in spite of the
appearance of $|\l_2\l_3|$ in the denominator; see \cite{awe2}.)
The operators $e^{i\bar\mu_2 c_2}$ and $e^{i\bar\mu_3c_3}$ have
analogous action.

We are now ready to write the Hamiltonian constraint explicitly in
the $\l_i$-representation. As noted above, the three terms in the
first square bracket on the right hand side of Eq. (\ref{qHam1})
constitute the gravitational part of $\hat{\mathcal{C}}_H$ for the
LQC of Bianchi I model%
\footnote{There are some minor changes in the action of these three
terms since $\gamma$ is no longer treated as a pseudoscalar (see
Appendix \ref{a1}), but these do not affect the discussion.}
and have been discussed in \cite{awe2}. In the next two
sub-sections we will now discuss the last two terms, which are
specific to the Bianchi II model.

\subsubsection{The Fourth term in $\hat{\mathcal{C}}_H$}
\label{s3.3.2}

Using a symmetric factor ordering, the fourth term becomes
\be \label{hc4} \hat{\mathcal{C}}_H^{(4)} = -\f{\alpha
p_2p_3\sqrt{|p_2p_3|}} {16\pi
G\gamma^2\sqrt\Delta\lp}\,\,\widehat{|p_1|^{-1/4}}\,(\hat{\ve}\,
\sin\bar\mu_1c_1+\sin\bar\mu_1c_1\,\hat{\ve})\,\widehat{|p_1|^{-1/4}}
\, . \ee
(Note that $p_2$ and $p_3$ commute with the other terms in
$\hat{\mathcal{C}}_H^{(4)}$). The operator $p_1$ is self-adjoint on
$\H_{\rm kin}^{\rm grav}$ whence any measurable function of $p_1$ is
also a well-defined self-adjoint operator. However, since kets
$|\l_1=0, \l_2,\l_3\rangle$ are normalizable in $\H_{\rm kin}^{\rm
grav}$, the naive inverse powers of $\hat{p}_1$ fail to be densely
defined and cannot be self-adjoint. To define inverse powers, as is
usual in LQG, we will use a variation on the Thiemann inverse triad
identities \cite{tt}. Classically, we have the identity
\be \label{class} |p_1|^{-1/4} = -\f{i\,\sgn(p_1)}{2\pi G\gamma}
\sqrt\f{|p_2p_3|}{\Delta\lp^2}\,\,
e^{-i\bar\mu_1c_1}\,\{e^{i\bar\mu_1c_1},|p_1|^{1/4}\}\, . \ee
which holds for any choice of $\bar\mu_1$. Since it is most natural
to use the same $\bar\mu_1$ that featured in the definition of the
connection operator, we will make this choice. Eq (\ref{class})
suggests a natural quantization strategy for $|p_1|^{-1/4}$. Using
it and the parity considerations, we are led to the following factor
ordering:%
\footnote{In the classical theory, $(L_2L_3)^{1/4}\,|p_1|^{-1/4}$ is
independent of the choice of the elementary cell. As pointed out in
\cite{kv1} the inverse triad operators, by contrast, depend on the
choice of the cell. However, one can verify that as we remove the
regulator, i.e., take the limit $\mathcal{V} \to \R^3$, as in the
classical theory, the limiting
$(L_2L_3)^{1/4}\,\widehat{|p_1|^{-1/4}}$ has a well defined limit.}
\be \widehat{|p_1|^{-1/4}} = - \f{i\,\sgn(p_1)}{2\pi
G\gamma}\sqrt\f{|p_2p_3|}{\Delta\lp^2}\,\,e^{-i\bar\mu_1c_1/2}\,\,
\f{1}{i\hbar}[e^{i\bar\mu_1c_1},|p_1|^{1/4}]\,\,
e^{-i\bar\mu_1c_1/2}\, , \ee
where, as is common in LQC, $\sgn(p_1)$ is defined as
\be \sgn(p_1) = \left\{\rlap{\rlap{\raise4ex\hbox{\,\,$+1$ if $p_1>0$,}}
{\raise0ex\hbox{\,\,$0$ if $p_1=0$,}}}
{\lower4ex\hbox{\,\,$-1$ if $p_1<0$.}} \right. \ee

At first it may seem surprising that the expression of
$\widehat{|p_1|^{-1/4}}$ involves operators other than ${p_1}$. It
is therefore important to verify that it has the standard desirable
properties. First, as one would hope, it is indeed diagonal in the
eigenbasis of the operators $\hat{p}_k$:
\be \label{inv} \widehat{|p_1|^{-1/4}}\, |\l_1,\l_2,\l_3\rangle =
\f{\sqrt2 \sgn(\l_1)\,\sqrt{|\l_2\l_3|}}
{(4\pi\gamma\sqrt\Delta\lp^3)^{1/6}}
\left(\sqrt{|v+\sgn(\l_2\l_3)|}-\sqrt{|v-\sgn(\l_2\l_3)|}\right)\,
|\l_1,\l_2,\l_3\rangle. \ee
Second, on eigenkets with large volume, the eigenvalue is indeed
well-approximated by $p_1^{-1/4}$, whence on semi-classical states
it behaves as the inverse of $\hat{p}^{1/4}$, just as one would
hope. Thus, (\ref{inv}) is a viable candidate for
$\widehat{|p_1|^{-1/4}}$. But there are interesting
non-trivialities in the Planck regime. In particular, although
counter-intuitive, as is common in LQC the operator annihilates
states $|\l_1,\l_2,\l_3\rangle$ with $v = 2\l_1\l_2\l_3 =0$

Finally, note that the operator $\hat{\ve}$ appearing in the
expression (\ref{hc4}) of $\hat{\mathcal{C}}_H^{(4)}$ either
operates immediately before or after $\widehat{|p_1|^{-1/4}}$. Since
$\widehat{|p_1|^{-1/4}}$ annihilates all zero volume states and
$\hat{\ve}$ acts on such states as the identity operator, we only
need to consider the action of $\hat{\ve}$ on states with nonzero
volume. In this case, $\hat{\ve}$ acts as $\sgn(v)$. Therefore the
action of $\hat{\mathcal{C}}_H^{(4)}$ can be written as:
\begin{align}
\Big[\hat{\mathcal{C}}_H^{(4)}\,\Psi\Big](\l_1,\l_2,\l_3) =&
-\f{i\alpha\pi\sqrt\Delta\hbar\lp^2}{(4\pi\gamma\sqrt\Delta)^{1/3}}\,\,
\sgn(v)\,\, (\l_2\l_3)^4\nonumber\\
\left(\sqrt{|v+\sgn(\l_2\l_3)|}-\sqrt{|v-\sgn(\l_2\l_3)|} \right)
& \quad \Big[\Phi^+(\l_1,\l_2,\l_3) -\Phi^-(\l_1,\l_2,\l_3)\Big]
\label{c4}
\end{align}
where
\begin{align} \Phi^\pm(\l_1,\l_2,\l_3) =& \Big(\sqrt{\left|v\pm2\sgn(\l_2
\l_3)+\sgn(\l_2\l_3))\right|} 
-\sqrt{\left|v\pm2\sgn(\l_2\l_3)-\sgn(\l_2\l_3)\right|}\, \Big)
\nonumber \\  & \quad\: \times
\big(\sgn(v)+\sgn(v\pm2 \sgn(\l_2\l_3))\big)\,\,
\Psi(\f{v\pm2\sgn(\l_2\l_3)}{v}\l_1,\l_2,\l_3).\label{phi} \end{align}

Recall that in the classical theory the singularity corresponds
precisely to the phase space points at which the volume vanishes.
Therefore, as in the Bianchi I model, states with support only on
points with $v=0$ will be called `singular' and those which vanish
at points with $v=0$ will be called regular. The total Hilbert space
$\Hkg$ is naturally decomposed as a direct sum $\Hkg = \H^{\rm
grav}_{\rm sing}\oplus \H^{\rm grav}_{\rm reg}$ of singular and
regular sub-spaces. We will conclude this discussion by examining
the action of $\hat{\mathcal{C}}_H^{(4)}$ on these sub-spaces. Note
first that in the action (\ref{hc4}) of $\hat{\mathcal{C}}_H^{(4)}$,
the state is first acted upon by the operator
$\widehat{|p_1|^{-1/4}}$. Since this operator annihilates states
$|\l_1 \l_2,\l_3\rangle$ with $v = 2\l_1\l_2\l_3 =0$, singular
states are simply annihilated by $\hat{\mathcal{C}}_H^{(4)}$. In
particular this implies that the singular sub-space is mapped to
itself under this action. It is clear from (\ref{phi}) that if
$\Psi$ is regular, i.e. vanishes on all points with $v =0$,
$\Phi^\pm$ also vanish at these points. Thus the regular sub-space
is also preserved by this action. This fact will be used in the
discussion of singularity resolution in section \ref{s3.3.4}.

\emph{Remark:}\, Our definition of the operator
$\widehat{|p|^{-1/4}}$ is not unique; as is common with non-trivial
functions of elementary variables, there are factor ordering
ambiguities. For example, for $0<n<1/2$, we have the classical
identity
\be
|p_1|^{n-1/2}=\f{-i\sgn(p_1)\sqrt{|p_2p_3|}}{8\pi\gamma\sqrt\Delta
G\lp n} e^{-i\bar\mu_1c_1}\left\{e^{i\bar\mu_1c_1},|p_1|^n\right\}\,
. \nonumber \ee
Hence, it is possible to instead define $\widehat{p_1^{-1/4}}$ as
\be \widehat{p_1^{-1/4}} =
\left(\widehat{|p_1|^{n-1/2}}\right)^{-1/(4n-2)} \nonumber \ee
where
\be \widehat{|p_1|^{n-1/2}} =
-\f{(4\pi\gamma\sqrt\Delta\lp^3)^{(2+2n)/3}} {4^n(8\pi\gamma
G\sqrt\Delta\lp)^3n}\sgn(\l_1)|\l_2\l_3|^{1-2n}\Big[|v+
\sgn(\l_2\l_3)|^{2n}-|v-\sgn(\l_2\l_3)|^{2n}\Big]. \nonumber \ee
For $n\ne1/4$, this choice for the operator $\widehat{p_1^{-1/4}}$
is not equivalent to the one we chose.  These two choices are both
well-defined and admit the same classical limit but they differ in
the Planck regime. It is also possible to construct other such
inequivalent $\widehat{p_1^{-1/4}}$ candidate operators. For
definiteness we have made the `simplest' choice.

\subsubsection{The fifth term in $\hat{\mathcal{C}}_H$}
\label{s3.3.3}

Let us now consider the last term in the expression of the
gravitational part of the Hamiltonian constraint
\be \hat{\mathcal{C}}_H^{(5)} = \f{\alpha^2}{32\pi
G\gamma^2}(1+\gamma^2)\,(p_2 p_3)^2\,\,\widehat{p_1^{-2}}. \ee
This term is simpler since it only involves powers of $p_k$ and we
are working in a representation where $p_k$ are diagonal. From our
discussion of the last section, it is natural to set
\be \widehat{p_1^{-2}}:=\left(\widehat{p_1^{-1/4}}\right)^8\, , \ee
then we have
\begin{align} \label{c5} \hat{\mathcal{C}}_H^{(5)}\,\Psi(\l_1,\l_2,
\l_3)\, =\, & \f{8\pi\alpha^2\Delta(1+\gamma^2)\hbar\lp^2}{(4\pi\gamma\sqrt
\Delta)^{2/3}}\,\sgn(\l_1)^8\l_2^8\l_3^8 \nonumber \\ & \times \left(
\sqrt{|v+\sgn(\l_2\l_3)|}-\sqrt{|v-\sgn(\l_2\l_3)|}\right)^8 \Psi(\l_1,\l_2,
\l_3). \end{align}
Again, it is clear that if $v=0$, the wave function is annihilated
by this part of the constraint.  Also, it follows by inspection that
the singular and regular subspaces are both mapped to themselves by
the action of $\hat{\mathcal{C}}_H^{(5)}$.

\subsubsection{Singularity resolution}
\label{s3.3.4}

We can now determine the gravitational part $\hat{\mathcal{C}}_\g$
of the Hamiltonian constraint by combining the results of
\cite{awe2} and Eqs. (\ref{c4}) and (\ref{c5}). We have:
\be \label{qHam2} \hat{\mathcal{C}}_\g = \hat{\mathcal{C}}_\g^{\rm
(BI)} + \hat{\mathcal{C}}_H^{(4)} + \hat{\mathcal{C}}_H^{(5)} \ee
where $\hat{\mathcal{C}}_\g^{\rm (BI)}$ is the gravitational part of
the Hamiltonian constraint in the Bianchi I model \cite{awe2}. There
is however, a conceptual subtlety. In the classical theory the
Hamiltonian density $\mathcal{C}_\g/(L_1L_2L_3)^2$ is independent of
the choice of the elementary cell (where we have to divide by
$(L_1L_2L_3)^2$ because the lapse corresponding to harmonic time
scales as $(L_1L_2L_3)$ and the Hamiltonian constraint is obtained
by integration over the elementary cell $\mathcal{V}$). As shown in
the section V of \cite{awe2}, $\hat{\mathcal{C}}_\g^{\rm
(BI)}/(L_1L_2L_3)^2$ is again independent of the choice of the
elementary cell $\mathcal{V}$. However, the two additional terms
that are special to the Bianchi II model are not independent of
$\mathcal{V}$ because they involve the inverse-triad operators
\cite{kv1}. Nonetheless, in the limit as we take the regulator away,
i.e., $\mathcal{V} \to \R^3$, the operator
$\hat{\mathcal{C}}_\g/(L_1L_2L_3)^2$ has a well-defined limit (see
footnote 5). Strictly speaking, in the discussion of Bianchi II
quantum dynamics, we have to work with this limit, rather than with
operators defined using a fixed cell.

As in the Bianchi I model, the action simplifies if we replace one
of the $\l_i$ by $v$.  In the Bianchi I model, it does not matter
which of the $\l_i$ is replaced because of the additional symmetry
of that model. In the Bianchi II case, while it remains possible to
replace any of the $\l_i$, it is simplest to replace $\l_1$ by $v$
and represent quantum states as $\Psi=\Psi(\l_2,\l_3,v;T)$. This
change of variables would be nontrivial if, as in the Wheeler-DeWitt
theory, we had used the Lesbegue measure in the gravitational
sector. However, it is quite tame here because the norms are defined
using a discrete measure. The inner product on $\Hkg$ is now given
by
\be \langle\Psi_1|\Psi_2\rangle_{\rm kin} = \sum_{\l_2,\l_3,v}
\,\,\bar{\Psi}_1(\l_2,\l_3,v)\,\Psi_2(\l_2,\l_3,v) \ee
and states are symmetric under the action of $\hat\Pi_k$. In
Appendix \ref{a1}, we show that, under the action of reflections
$\hat\Pi_i$, the operators $\sin\bar\mu_ic_i$ have the same
transformation properties that $c_i$ have under reflections $\Pi_i$
in the classical theory. As a consequence, $\hat{\mathcal{C}}_\g$ is
also reflection symmetric. Therefore, its action is well defined on
$\Hkg$: $\hat{\mathcal{C}}_\g$ is a densely defined, symmetric
operator on this Hilbert space. In the isotropic case, its analog
has been shown to be essentially self-adjoint \cite{warsaw2}. In
what follows we will assume that (\ref{qHam2}) is essentially
self-adjoint on $\Hkg$ and work with its self-adjoint extension.

It is now straightforward to write down the full Hamiltonian
constraint on $\Hkg$:
\be \label{qHam3} -\hbar^2\, \partial^2_T \, \Psi(\l_2,\l_3,v; T) =
\Theta\, \Psi(\l_2,\l_3,v; T)\quad {\rm where}\quad \Theta = -2
\hat{\mathcal{C}}_\g \ee
As in the isotropic case \cite{aps2}, one can obtain the physical
Hilbert space $\H_{\rm phy}$ by a group averaging procedure and the
final result is completely analogous. Elements of $\H_{\rm phy}$
consist of `positive frequency' solutions to (\ref{qHam3}), i.e.,
solutions to
\be \label{qHam4} -i\hbar \partial_T \Psi(\l_2,\l_3,v; T)\, = \,
\sqrt{|\Theta|}\, \Psi(\l_2,\l_3,v; T)\, ,\ee
which are symmetric under the three reflection maps $\hat\Pi_i$,
i.e. satisfy
\be \label{sym} \Psi(\l_2,\l_3,v;\, T) = \Psi(|\l_2|,|\l_3|,|v|;\,
T)\, . \ee
The scalar product is given simply by:
\ba \label{ip1} \langle \Psi_1|\Psi_2\rangle_{\rm phys} &=& \langle
\Psi_1(\l_2,\l_3, v; T_o)|\Psi_2(\l_2,\l_3,v; T_o) \rangle_{\rm
kin} \nonumber\\
&=& \sum_{\l_1,\l_2,\l_3} \bar\Psi_1(\vec{\l}, T_o)\,
\Psi_2(\vec{\l}, T_o) \ea
where $T_o$ is any ``instant'' of internal time $T$.

We can now address the issue of singularity resolution using general
properties of various operators. Recall that the gravitational part
of the Hamiltonian constraint operator in the Bianchi I model shares
two properties with the fourth and the fifth terms studied above
which are specific to the Bianchi II model. First, it annihilates
singular states and, second, singular states decouple from the
regular states under its action. Therefore the full Bianchi II
Hamiltonian constraint also has these two properties. Since the
singular states decouple from regular states%
\footnote{Singular states are in the kernel of $\Theta$ and regular
states are orthogonal to the singular ones. From spectral
decomposition one expects $\sqrt{\Theta}$ to have the same property.
However, to complete this argument, one would have to establish that
$\hat{\mathcal{C}}_\g$ is essentially self-adjoint and its self
adjoint extension also shares this property.},
an initial state in the regular sub-space cannot become singular
during evolution. It is in this precise sense that the classical
singularity is resolved. Sometimes one considers weaker forms of
singularity resolution. For example, it could happen that the
evolution of the wave function is always well defined but a regular
state can evolve to the singular sub-space. For the Bianchi I and II
models, the singularity is resolved in a stronger sense: \emph{Not
only is the evolution well defined at all times, but the singular
states (are stationary and) decouple entirely from the regular
ones.}

\subsubsection{The explicit form of the Hamiltonian constraint}
\label{s.3.5}

We will conclude by providing an explicit form of the full quantum
constraint equation that will be needed in numerical simulations.

Recall that in the Bianchi I model \cite{awe2} symmetries enabled us
to restrict our attention to the positive octant of the
3-dimensional space spanned by $(\l_1,\l_2,\l_3)$. This is again the
case for the Bianchi II model. More precisely, elements of $\Hkg$
are invariant under the three parity maps $\hat{\Pi}_k$ and, as
shown in the Appendix \ref{a1}, the Hamiltonian constraint satisfies
$\hat{\Pi}_k\, \hat{\mathcal{C}}_\g \hat{\Pi}_k =
\hat{\mathcal{C}}_\g$. Therefore, knowledge of the restriction of
the image $\hat{ \mathcal{C}}_\g\Psi$ of $\Psi$ to the positive
octant suffices to determine $\hat{\mathcal{C}}_\g\Psi$ completely.
In the positive octant, $\sgn(\l_k)$ can only be 0 or 1 which
simplifies the action of operators. Therefore, in the remainder of
this section we will restrict the argument of $\hat{
\mathcal{C}}_H\Psi$ to the positive octant. The full action is given
simply by
\be \big(\hat{\mathcal{C}}_\g\Psi\big)(\l_2,\l_3, v)=\big(\hat{
\mathcal{C}}_\g\Psi\big)(|\l_2|,|\l_3|,|v|). \ee

Since the singular states are annihilated by
$\hat{\mathcal{C}}_\g$, their evolution is trivial:
\be \partial_T^2\, \Psi(\l_2,\l_3,v=0;T) = 0\, . \ee
Non-singular states are physically more relevant. On them, the
explicit form of the full constraint is given by:
\begin{align} \partial_T^2\, \Psi(\l_2,\l_3,v;T) =& \f{\pi G}{2}\Bigg[\sqrt{v}
\bigg((v+2)\sqrt{v+4}\,\Psi^+_4(\l_2,\l_3,v;T) - (v+2)\sqrt v\,
\Psi^+_0( \l_2, \l_3,v;T)\nonumber \\& -(v-2)\sqrt
v\,\Psi^-_0(\l_2,\l_3,v;T)+(v-2)\sqrt{|v-4|}
\,\Psi^-_4(\l_2,\l_3,v;T)\bigg)\nonumber \\ &
+\f{2i\alpha\sqrt\Delta}{(4\pi
\gamma\sqrt\Delta)^{1/3}}(\l_2\l_3)^4\left(\sqrt{v+1}-\sqrt{|v-1|}\right)\bigg(
\Phi^--\Phi^+\bigg)(\l_2,\l_3,v;T) \nonumber\\& \label{qHamfin} +
\f{16
\alpha^2\Delta(1+\gamma^2)}{(4\pi\gamma\sqrt\Delta)^{2/3}}(\l_2\l_3)^8\:
(\sqrt{v+1}-\sqrt{|v-1|})^8\:\Psi(\l_2,\l_3,v;T)\Bigg] \end{align}
where $\Psi^\pm_{0,4}$ are defined as follows:
\begin{align} \Psi^\pm_4(\l_2,\l_3,v;T)=& \:\Psi\left(\f{v\pm4}{v\pm2}\cdot
\l_2,\f{v\pm2}{v}\cdot\l_3,v\pm4;T\right)+\Psi\left(\f{v\pm4}{v\pm2}\cdot\l_2,
\l_3,v\pm4;T\right)\nonumber\\& +\Psi\left(\f{v\pm2}{v}\cdot\l_2,\f{v\pm4}{v
\pm2}\cdot\l_3,v\pm4;T\right)+\Psi\left(\f{v\pm2}{v}\cdot\l_2, \l_3,v\pm4;T
\right) \nonumber \\ & +\Psi\left(\l_2,\f{v\pm2}{v}\cdot\l_3,v\pm4;T\right)+
\Psi\left(\l_2,\f{v\pm4}{v\pm2}\cdot\l_3,v\pm4;T\right), \end{align}
and
\begin{align} \Psi^\pm_0(\l_2,\l_3,v;T)= & \:\Psi\left(\f{v\pm2}{v}\cdot\l_2,
\f{v}{v\pm2}\cdot\l_3,v;T\right)+\Psi\left(\f{v\pm2}{v}\cdot\l_2,\l_3,v;T
\right) \nonumber \\ & +\Psi\left(\f{v}{v\pm2}\cdot\l_2,\f{v\pm2}{v}\cdot\l_3,
v;T\right)+\Psi\left(\f{v}{v\pm2}\cdot\l_2,\l_3,v;T\right) \nonumber \\ & +
\Psi\left(\l_2,\f{v}{v\pm2}\cdot\l_3,v;T\right)+\Psi\left(\l_2,\f{v\pm2}{v}
\cdot\l_3,v;T\right)\, ,\end{align}
while $(\Phi^--\Phi^+)$ is given by
\begin{align} \big(\Phi^--\Phi^+\big)(\l_2,\l_3,v;T)\, =& \,(\sqrt{|v-2+
\sgn(v-2)|}-\sqrt{|v-2-\sgn(v-2)|}) \nonumber \\ & \qquad \times (1+\sgn(v-2))
\Psi(\l_2,\l_3,v-2;T) \nonumber \\ & \: -2(\sqrt{v+3}-\sqrt{v+1})
\Psi(\l_2,\l_3,v+2;T). \end{align}
(The imaginary coefficients in (\ref{qHamfin}) come from the
action of single $\sin\bar\mu_ic_i$ terms.)

Eq. (\ref{qHamfin}) immediately implies that, as in the Bianchi I
model, the steps in $v$ are uniform: the argument of the wave
function only involves $v-4, v-2, v, v+2$ and $v+4$. Thus, there is
a superselection in $v$. For each $\epsilon\in[0,2),$ let us
introduce a lattice $\mathcal{L}_\epsilon$ of points $v=2n+\epsilon$
if $\epsilon$ is 0 or 1 or $v=n+\epsilon$ otherwise.
\footnote{The lattice for $\epsilon \not= 0,1$ is twice as large as
that for $\epsilon=0$ or $\epsilon=1$ due to the symmetry properties
of the wave function.}
Then the quantum evolution ---and the action of the Dirac
observables $\hat{p}_T$ and $\hat{V}|_{T}$ commonly used in
LQC--- preserves the subspaces $\mathcal{H}^\epsilon_{\mathrm{phy}}$
consisting of states with support in $v$ on $\mathcal{L}_\epsilon$.
The most interesting lattice is the one corresponding to
$\epsilon=0$
since it includes the classically singular points $v=0$. 

Finally, it is obvious from (\ref{qHamfin}) that in the limit
$\alpha\to0$ quantum dynamics of the Bianchi II model reduces to
that of the Bianchi I model discussed in \cite{awe2}. In particular,
it is possible to obtain the LQC dynamics for the $k$=0 FRW
cosmology from this model by first taking $\alpha\to0$ and then
following the projection map defined in section IVA in \cite{awe2}.

\section{Effective Equations}
\label{s4}

In the isotropic models, effective equations have been introduced
via two different approaches ---the embedding and the truncation
methods. Both start by regarding the space of quantum states as an
infinite dimensional symplectic manifold ---the quantum phase
space--- which is also equipped with a K\"ahler structure that
descends from the Hermitian inner product on the Hilbert space. In
the first method, one finds a judicious embedding of the classical
phase space into the quantum phase space which is approximately
preserved by the quantum evolution vector field \cite{jw,vt}. By
projecting this vector field into the image of the embedding one
obtains quantum corrected effective equations. In the isotropic case
these effective equations provide an excellent approximation to the
full quantum evolution of states which are Gaussians at late times,
even in the $\Lambda\not=0$ as well as k=$\pm 1$ cases where the
models are not exactly soluble. In the second method one uses
expectation values, uncertainties, and higher moments to define a
convenient system of coordinates on the infinite dimensional phase
space. The exact quantum evolution equations are then a set of
coupled non-linear ordinary differential equations for these
coordinates. By a judicious truncation of this system one obtains
effective equations containing quantum corrections \cite{bs}. In its
spirit the first method is analogous to the `variational principle
technique' used in perturbation theory, in that it requires a
judicious combination of art (of selecting the embedding) and
science. It is often simpler to use and can be surprisingly
accurate. The second method is more systematic, similar in our
analogy to the standard, order by order perturbation theory. It is
also more general in the sense that it is applicable to a wide
variety of states. In this section we will use the first method to
gain qualitative insights into leading order quantum effects.

To obtain the effective equations, without loss of generality we can
restrict our attention to the positive octant of the classical phase
space (where $\ve=1$). Then the quantum corrected Hamiltonian
constraint is given by the classical analogue of (\ref{qHam1}):
\be \label{Heff}
\f{p_T^2}{2}+\mathcal{C}^{\mathrm{eff}}_{\mathrm{grav}} = 0, \ee
where
\begin{align} \mathcal{C}^{\mathrm{eff}}_{\mathrm{grav}} =&
-\f{p_1p_2p_3}{8\pi G\gamma^2\Delta\lp^2}
\Bigg[\sin\bar\mu_1c_1\sin\bar\mu_2c_2+\sin\bar\mu_2c_2
\sin\bar\mu_3c_3+\sin\bar\mu_3c_3\sin\bar\mu_1c_1\Bigg] \nonumber
\\ & \quad - \f{1}{8\pi
G\gamma^2}\Bigg[\f{\alpha(p_2p_3)^{3/2}}{\sqrt\Delta\lp
\sqrt{p_1}}\sin\bar\mu_1c_1 -(1+\gamma^2)\left(\f{\alpha
p_2p_3}{2p_1}\right)^2 \Bigg]. \end{align}
Using the expressions (\ref{mubar}) of $\bar\mu_k$, it is easy to
verify that far away from the classical singularity ---more
precisely in the regime in which the (gauge fixed) spin connection
and the extrinsic curvature are sufficiently small so that
$c_k\bar\mu_k \ll 1$---  the effective Hamiltonian constraint
(\ref{Heff}) is well-approximated by the classical one (\ref{Hcl}).

Since $\sin\theta$ is bounded by 1 for all $\theta$, these equations
imply that the matter density $\rho_{\mathrm{matt}}=
p_T^2/2V^2=p_T^2/2p_1p_2p_3$ satisfies
\be \rho_{\mathrm{matt}} \le \f{3}{8\pi\gamma^2\Delta G\lp^2}+
\f{1}{8\pi \gamma^2 G} \left[\f{x}{\sqrt\Delta\lp} -
\f{(1+\gamma^2)x^2}{4} \right] \ee
where we have introduced $x:=\alpha\sqrt{p_2p_3/p_1^3}$. The maximum
of the expression in square brackets is attained at
$x=2/(1+\gamma^2)\sqrt\Delta\lp$, whence
\be \rho_{\mathrm{matt}} \le \f{3+(1+\gamma^2)^{-1}}
{8\pi\gamma^2\Delta G \lp^2} \approx 0.54 \rho_{\mathrm{Pl}}. \ee
Thus, on the constraint surface in the phase space defined by
(\ref{Heff}), the matter energy density is bounded by $0.54
\rho_{\mathrm{Pl}}$. But this bound may be far from being optimal.
In all isotropic models, the optimal bound on matter density was
found to be $0.41 \rho_{\rm Pl}$. In the Bianchi I model, available
simulations by Vandersloot (private communication) show that the
`volume bounce' occurs when matter density is \emph{lower} than
$0.41 \rho_{\rm Pl}$ because there is also energy density in
gravitational waves. It would be interesting to use numerical
simulations to find out what happens for generic solutions to the
Bianchi II effective equations.

Finally, to obtain the effective equations for each variable, one
simply takes its Poisson bracket with the effective Hamiltonian
constraint.  This gives the effective equations
\be \dot{p_1} = \gamma^{-1}\left(\f{p_1^2}{\bar\mu_1}(\sin\bar\mu_2c_2+
\sin\bar\mu_3c_3)+\alpha p_2p_3\right)\cos\bar\mu_1c_1, \ee
\be \dot{p_2} = \f{p_2^2}{\gamma\bar\mu_2}(\sin\bar\mu_1c_1+\sin\bar\mu_3c_3)
\cos\bar\mu_2c_2, \ee
\be \dot{p_3} = \f{p_3^2}{\gamma\bar\mu_3}(\sin\bar\mu_1c_1+\sin\bar\mu_2c_2)
\cos\bar\mu_3c_3, \ee
\begin{align} \dot{c_1} &= -\f{1}{\gamma}\Big[\f{p_2p_3}{\Delta\lp^2}\big(
\sin\bar\mu_1c_1\sin\bar\mu_2c_2+\sin\bar\mu_1c_1\sin\bar\mu_3c_3+\sin\bar\mu_2
c_2\sin\bar\mu_3c_3 \nonumber \\ & \qquad +\f{\bar\mu_1c_1}{2}\cos\bar\mu_1c_1(
\sin\bar\mu_2c_2+\sin\bar\mu_3c_3)-\f{\bar\mu_2c_2}{2}\cos\bar\mu_2c_2(\sin\bar
\mu_1c_1+\sin\bar\mu_3c_3) \nonumber \\ & \qquad -\f{\bar\mu_3c_3}{2}\cos\bar
\mu_3c_3(\sin\bar\mu_1c_1+\sin\bar\mu_2c_2)\big)+(1+\gamma^2)\alpha^2\f{(p_2
p_3)^2}{2p_1^3} \nonumber \\ & \qquad +\f{\alpha}{2\sqrt\Delta\lp}\left(
\f{p_2p_3}{p_1}\right)^{3/2}\!\!(\bar\mu_1c_1\cos\bar\mu_1c_1-\sin\bar\mu_1c_1)
\Big], \end{align}
\begin{align} \dot{c_2} &= -\f{1}{\gamma}\Big[\f{p_1p_3}{\Delta\lp^2}\big(
\sin\bar\mu_1c_1\sin\bar\mu_2c_2+\sin\bar\mu_1c_1\sin\bar\mu_3c_3+\sin\bar\mu_2
c_2\sin\bar\mu_3c_3 \nonumber \\ & \qquad -\f{\bar\mu_1c_1}{2}\cos\bar\mu_1c_1(
\sin\bar\mu_2c_2+\sin\bar\mu_3c_3)+\f{\bar\mu_2c_2}{2}\cos\bar\mu_2c_2(\sin\bar
\mu_1c_1+\sin\bar\mu_3c_3) \nonumber \\ & \qquad -\f{\bar\mu_3c_3}{2}\cos\bar
\mu_3c_3(\sin\bar\mu_1c_1+\sin\bar\mu_2c_2)\big)-(1+\gamma^2)\alpha^2\f{p_2
p_3^2}{2p_1^2} \nonumber \\ & \qquad +\f{\alpha p_3}{2\bar\mu_1}(3\sin\bar
\mu_1c_1-\bar\mu_1c_1\cos\bar\mu_1c_1) \Big], \end{align}
\begin{align} \dot{c_3} &= -\f{1}{\gamma}\Big[\f{p_1p_2}{\Delta\lp^2}\big(
\sin\bar\mu_1c_1\sin\bar\mu_2c_2+\sin\bar\mu_1c_1\sin\bar\mu_3c_3+\sin\bar\mu_2
c_2\sin\bar\mu_3c_3 \nonumber \\ & \qquad -\f{\bar\mu_1c_1}{2}\cos\bar\mu_1c_1(
\sin\bar\mu_2c_2+\sin\bar\mu_3c_3)-\f{\bar\mu_2c_2}{2}\cos\bar\mu_2c_2(\sin\bar
\mu_1c_1+\sin\bar\mu_3c_3) \nonumber \\ & \qquad +\f{\bar\mu_3c_3}{2}\cos\bar
\mu_3c_3(\sin\bar\mu_1c_1+\sin\bar\mu_2c_2)\big)-(1+\gamma^2)\alpha^2\f{p_2^2
p_3}{2p_1^2} \nonumber \\ & \qquad +\f{\alpha p_2}{2\bar\mu_1}(3\sin\bar
\mu_1c_1-\bar\mu_1c_1\cos\bar\mu_1c_1) \Big]. \end{align}
In the ``embedding approach'' these effective equations provide the
leading-order quantum corrections to the classical equations of
motion Eqs.~(\ref{ceom1})~--~(\ref{ceom2}). It would be very
interesting to numerically test if the accuracy they display in the
isotropic case for states which are Gaussians at late times carries
over to the Bianchi II case.

\section{Discussion}
\label{s5}

In this paper, we analyzed the ``improved'' LQC dynamics of the
Bianchi II model.  As in the isotropic and Bianchi I cases, we chose
the matter source to be a massless scalar field since it continues
to serve as a viable relational time parameter in the classical as
well as the quantum theory. It is again rather straightforward to
accommodate additional matter fields in this framework.

Our broad strategy is the same as that used in the Bianchi I model
\cite{awe2}. However, because Bianchi II models have anisotropies
\emph{as well as} spatial curvature, holonomies around closed curves
are no longer guaranteed to be almost periodic functions of the
connection. Hence, one cannot use them to construct the field
strength operator on the LQC Hilbert space; a new conceptual and
technical input is necessary to define the quantum Hamiltonian
constraint operator. We overcame this difficulty by generalizing the
strategy used so far \cite{abl,aps3,warsaw,apsv,kv1,ls,bp,awe2}.
Specifically, we used holonomies around open segments parallel to
the fiducial triads $\e^a_i$ to define a connection operator. This
strategy is also inspired by methods introduced by Thiemann in the
full theory \cite{tt}. However, because of gauge fixing LQC does not
enjoy the manifest diffeomorphism invariance of full LQG. As a
consequence, in LQC one needs a principle to fix the `length' of the
open segment along which holonomy is evaluated. We required that the
`length' be so chosen that the field strength operator constructed
from the resulting connection should agree with that constructed
from holonomies around closed loops whenever the second construction
is available. This guarantees that (apart from `tame' factor
ordering ambiguities) the new procedure reduces to the one used in
the LQC literature before. Moreover, the strategy of defining the
Hamiltonian constraint through this connection operator can be used
also in more general contexts. In particular, it enables one to
overcome a conceptual limitation of the otherwise complete treatment
of the isotropic, k=$-1$ model given in \cite{kv1,ls}. More
importantly, it extends to more general class A Bianchi models. A
systematic treatment of the Bianchi IX model along the lines of this
paper would be especially interesting.

There is a second ---but primarily technical--- difference from the
Bianchi I case: The Hamiltonian operator now contains inverse powers
of $p_1$. This was handled following a general method introduced by
Thiemann to define inverse triad operators in LQG \cite{tt}. As
usual, there is a factor ordering ambiguity. In the main discussion
we used the simplest operator which has the same symmetries with
respect to parity as its classical counterpart.

After addressing these two issues, we obtained a well defined
quantum Hamiltonian constraint and showed that the singularity in
Bianchi II models is resolved in the same precise sense as in the
FRW and Bianchi I models. The Kinematical Hilbert space $\Hkg$ can
be decomposed as $\Hkg = \H_{\rm sing}^{\rm grav} \oplus \H^{\rm
grav}_{\rm reg}$ where states in the singular subspace have support
only on configurations with zero volume, while those in the regular
sub-space have no support on these singular configurations.
\emph{The Hamiltonian constraint annihilates states in $\H_{\rm
sing}^{\rm grav}$ and maps $\H^{\rm grav}_{\rm reg}$ to itself.} We
also provided an explicit form of the Hamiltonian constraint which
should be helpful in performing numerical simulations.

Finally, we obtained effective equations using the ``embedding
method'' introduced by Willis \cite{jw} and further developed by
Taveras \cite{vt} in the isotropic case. There, although the
assumptions made in the derivation fail in the deep Planck regime,
the final equations provide a surprisingly accurate approximation to
the full quantum evolution of states which are Gaussians at late
times. This holds not only for the exactly soluble k=0, $\Lambda=0$
model but also for the much more complicated $\Lambda\not=0$ and
k=$\pm 1$ models. It would be interesting to see if this phenomenon
extends also to Bianchi II models. Furthermore, numerical solutions
of these effective equations themselves may be of considerable
interest because the simplest upper bound on matter density they
lead to is higher than that in all other models studied so far,
including Bianchi I. Numerical simulations of effective equations
will answer several questions within this approximation. Is the
upper bound optimal, i.e., do generic solutions to effective
equations come close to saturating it? In the Bianchi I case,
numerical simulations by Vandersloot (private communication)
revealed that, unlike in the isotropic model, there are several
distinct kinds of `bounces.' Roughly, anytime a shear ---or a Weyl
curvature--- scalar enters the Planck regime, quantum geometry
repulsion comes into pay in a dominant manner and `dilutes' that
scalar, preventing a blow up. How do additional terms in the Bianchi
II effective equations affect this scenario? Qualitative lessons
from numerical simulations would be valuable in developing further
intuition for various quantum geometry effects.

\section*{Acknowledgements:}

We would like to thank Gianluca Calcagni, Alex Corichi, Jerzy
Lewandowski, Simone Mercuri, Tomasz Pawlowski and Hanno Sahlmann for
helpful discussions. This research was supported in part by NSF
grant PHY0854743, the George A. and Margaret M. Downsbrough
Endowment, the Eberly research funds of Penn State, Le Fonds
qu\'eb\'ecois de la recherche sur la nature et les technologies and
the Edward A. and Rosemary A. Mebus Fellowship.

\begin{appendix}

\section{Parity Symmetries}
\label{a1}

In non-gravitational physics, parity transformations are normally
taken to be discrete diffeomorphisms $x_i \rightarrow -x_i$ in the
physical space which are isometries of the flat 3-metric thereon. In
the phase space formulation of general relativity,  we do not have a
flat metric ---or indeed, any fixed metric. Therefore these discrete
symmetries are no longer meaningful (except in the weak field
limit). However, if the dynamical variables have internal indices
---such as the triads and connections used in LQG--- we can use the
fact that the internal space $I$ is a vector space equipped with a
flat metric $q_{ij}$ to define parity operations on the internal
indices. Associated with any unit internal vector $v^i$, there is a
parity operator $\Pi_v$ which reflects the internal vectors across
the 2-plane orthogonal to $v^i$. This operation induces a natural
action on triads $e^a_i$, the connections $A_a^i$ and the conjugate
momenta $P^a_i =: (1/8\pi G\gamma)\, E^a_i$ (since they are internal
vectors or co-vectors).

The triads $e^a_i$ are proper internal co-vectors. In previous
references \cite{bd,awe2}, conventions were such that the spin
connection $\Gamma^a_i$ turned out to be an internal \emph{pseudo}
vector. It was then natural to regard the Barbero-Immirzi parameter
$\gamma$ to be a pseudo quantity so that the connection $A_a^i$ has
definite parity namely, it transforms as an internal pseudo-vector.
This in turn led to the conclusion that $P^a_i$ is also an internal
pseudo-vector (as one would expect because it is canonically
conjugate to $A_a^i$) \cite{awe2}. While this is all
self-consistent, these conventions lead to two undesirable
consequences. First, in the classical theory, it is not possible to
reconstruct the triads $e^a_i$ unambiguously starting from the
momenta $P^a_i$. Therefore, one cannot recover the space-time
geometry unambiguously starting from the Hamiltonian theory. Second,
the momenta $P^a_i$ are subject to a non-holonomic constraint which
obstructs the passage to quantum theory a la LQG. However, if one
sets conventions as in section \ref{s2.1}, then $\Gamma_a^i, \gamma,
A_a^i$ and $P_a^i$ are all \emph{proper} quantities and the two
difficulties disappear \cite{aa-dis}. In the main text we have used
this strategy. We now summarize the differences from the Appendix of
\cite{awe2} that it leads to.

In diagonal Bianchi models, we can restrict ourselves just to three
parity operations $\Pi_i$. Under their action, the canonical
variables $c_i,p_i$ transform as follows:
\be \label{P1} \Pi_1 (c_1,c_2, c_3) = (-c_1, c_2, c_3), \quad \quad
\Pi_1 (p_1, p_2, p_3) = (-p_1, p_2, p_3)\, , \ee
and the action of $\Pi_2, \Pi_3$ is given by cyclic permutations.
Thus, $c^i$ and $p_i$ are \emph{proper} internal vectors and
co-vectors. Under any of these maps $\Pi_i$, the symplectic
structure (\ref{pb2}), the Hamiltonian (\ref{Hcl}), and hence also
the Hamiltonian vector field, are left invariant. This is just as
one would expect because $\Pi_i$ are simply large gauge
transformations of the theory under which the physical metric
$q_{ab}$ and the extrinsic curvature $K_{ab}$ do not change. Also,
it is clear from the action of (\ref{P1}) that if one knows the
dynamical trajectories on the octant $p_i\ge 0$ of the phase space,
then dynamical trajectories on any other octant can be obtained just
by applying a suitable (combination of) $\Pi_i$. Therefore, in the
classical theory one can restrict one's attention just to the
positive octant.

Let us now turn to the quantum theory. We now have three operators
$\hat\Pi_i$. Their action on states is given by
\be \hat\Pi_1 \Psi(\l_1,\l_2,\l_3) = \Psi(-\l_1,\l_2\l_3)\, , \ee
etc. What is the induced action on operators? Since
\begin{align} \label{P2}
\hat{\Pi}_1\hat{\l}_1\hat{\Pi}_1\Psi(\l_1,\l_2,\l_3)
&= \hat{\Pi}_1 \Big({\l}_1\,\Psi(-\l_1,\l_2,\l_3)\Big) \nonumber \\
&= -\l_1\Psi(\l_1,\l_2,\l_3), \end{align}
we have
\be \label{P3} \hat{\Pi}_1\hat{\l}_1\hat{\Pi}_1  = -\hat{\l}_1. \ee
The Hamiltonian constraint operator, modulo factor ordering which is
not important here, is given by Eq. (\ref{qHam1}). To calculate its
transformation property under parity maps, in addition to
(\ref{P3}), we also need the transformation property of the
operators $\sin \bar\mu_ic_i$ and $\hat{\ve}$ and operators
corresponding to inverse powers of $p_1$. Due to the symmetries of
type A Bianchi models, to know the properties of $\sin \bar\mu_ic_i$
under parity transformations, it is sufficient to calculate
$\hat{\Pi}_i \sin \bar\mu_1c_1\hat{\Pi}_i$.  We have:
\begin{align} \hat{\Pi}_1\sin\bar\mu_1c_1\hat{\Pi}_1\Psi(\l_1,\l_2,\l_3)
&= \f{1}{2i}\, \hat\Pi_1\,\Big[\Psi(-\l_1+\f{1}{|\l_2\l_3|},\l_2,\l_3)-
\Psi(-\l_1-\f{1} {|\l_2\l_3|},\l_2,\l_3)\Big] \nonumber \\
&= \f{1}{2i}\Big[\Psi(\l_1+\f{1}{|\l_2\l_3|},\l_2,\l_3)-\Psi(\l_1-
\f{1}{|\l_2\l_3|},\l_2,\l_3)\Big] \nonumber \\ &=
-\sin\bar\mu_1c_1\Psi(\l_1,\l_2,\l_3), \end{align}
whence
\be \hat{\Pi}_1 \sin\bar\mu_1c_1\hat\Pi_1 = -\sin\bar\mu_1c_1. \ee
An identical calculation shows that
\begin{align} \hat{\Pi}_2\sin\bar\mu_1c_1\hat{\Pi}_2\,\Psi(\l_1,\l_2,\l_3)
&= \f{1}{2i}\, \hat\Pi_2\, \Big[ \Psi(\l_1-\f{1}{|\l_2\l_3|},
-\l_2,\l_3)-\Psi(\l_1+\f{1}{|\l_2\l_3|},-\l_2,\l_3)\Big] \nonumber\\
&= \f{1}{2i}\Big[\Psi(\l_1-\f{1}{|\l_2\l_3|},\l_2,\l_3)-
\Psi(\l_1+\f{1}{|\l_2\l_3|},\l_2,\l_3)\Big] \nonumber \\
&= \sin\bar\mu_1c_1\Psi(\l_1,\l_2,\l_3)\, , \end{align}
and similarly for $\hat\Pi_3$. Therefore, we have:
\be \hat{\Pi}_2 \sin\bar\mu_1c_1\hat\Pi_2 = \sin\bar\mu_1c_1, \quad
{\rm and} \quad \hat{\Pi}_3 \sin\bar\mu_1c_1\hat{\Pi}_3 =
\sin\bar\mu_1c_1.\ee
As expected, these transformation properties of $\sin\bar\mu_1c_1$
under $\hat\Pi_i$ mirror those of $c_1$ under the three parity
operations $\Pi_i$ in the classical theory. (Note that, because of
the absolute value signs in the expressions (\ref{mubar}),
$\bar\mu_i$ do not change under any of the parity maps.)  Finally,
it is clear from Eq. (\ref{ve2}) that
\be  \hat\Pi_i\,\hat{\ve}\,\hat\Pi_i = \left\{ \rlap{\raise2ex\hbox{
$\hat{\ve}$ if $v=0$,}}{\lower2ex\hbox{$-\hat{\ve}$ otherwise,}}
\right. \ee
and from Eq. (\ref{inv}) that
\be \hat\Pi_i\,\widehat{|p_1|^{-1/4}}\,\hat\Pi_i =
\widehat{|p_1|^{-1/4}}. \ee
(Note incidentally that this need not be the case for different
factor ordering choices in Eq. (\ref{inv}).)

We can now collect these results to study the transformation
property of the Hamiltonian constraint. Consider first the regular
subspace $\H_{\rm reg}^{\rm grav}$ of $\Hkg$ spanned by states which
have no support on points with $v=0$. From Eq. (\ref{qHam1}) it
follows that the restriction to $\H_{\rm reg}^{\rm grav}$ of the
gravitational part of the Hamiltonian constraint is left invariant
under $\hat\Pi_i$. Since $\hat{p}_T^2$ is manifestly invariant, on
the regular sub-space we have
\be \label{Pham} \hat{\Pi}_i\,\, \hat{\mathcal{C}}_H \,\,\hat{\Pi}_i
= \hat{\mathcal{C}}_H \ee
Next, since the gravitational part of the Hamiltonian constraint
annihilates the states in the singular sub-space (i.e. those with
support only on those points at which $v=0$), we have
\be \hat{\mathcal{C}}_H\Psi=-\hbar^2\partial_T^2\Psi=\hat{\Pi}_i\,\,
\hat{\mathcal{C}}_H \,\, \hat{\Pi}_i\Psi. \ee
Thus, the Hamiltonian constraint operator is left invariant by all
the parity operators, mirroring the behavior of its classical
counterpart.

This invariance implies that, given any state $\Psi \in
\mathcal{H}_{\rm kin}^{\rm grav}$, the restriction to the positive
octant of its image under $\hat{\mathcal{C}}_{\rm grav}$ determines
its image everywhere on $\mathcal{H}_{\rm kin}^{\rm grav}$. This
property simplifies calculations and was used to arrive at the form
of the Hamiltonian constraint given in (\ref{qHamfin}).

\end{appendix}

\end{document}